\documentclass[aps,preprint,superscriptaddress]{revtex4}
\usepackage{bbm}
\usepackage{epsfig,amsopn}
\usepackage{graphicx}
\usepackage[T1]{fontenc}
\usepackage{amsmath,amssymb}
\usepackage{natbib}
\usepackage{footnote}
\usepackage[english]{babel}
\usepackage{pgf,pgfnodes}
\usepackage[latin1]{inputenc}
\renewcommand{\section}[1]{{\par\it #1.---}}
\def\be{\begin{equation}}
\def\ee{\end{equation}}
\def\bea{\begin{eqnarray}}
\def\eea{\end{eqnarray}}
\def\la{\langle}
\def\ra{\rangle}
\def\om{\omega}
\def\nn{\nonumber}

\def\f{\frac}

\begin{document}
\title{Non-equilibrium stationary state of a harmonic crystal with alternating masses} 
\author{Venkateshan Kannan}
\affiliation{Departments of Mathematics and Physics, 
    Rutgers University, Piscataway, NJ 08854}
\author{Abhishek Dhar}
\affiliation{Raman Research Institute, Bangalore 560080, India}
\author{J.L. Lebowitz$^1$}

\date{\today} 
\begin{abstract}
  We analyze the non-equilibrium steady states (NESS) of a one dimensional harmonic chain of $N$ atoms with alternating masses  connected to heat reservoirs 
at unequal temperatures. We find that the temperature profile defined through the local kinetic energy $T(j) \equiv {<p_j^2>}/{m_j}$, oscillates with period two in the bulk of the system. Depending on boundary conditions, either the 
heavier or the lighter  particles in the bulk are hotter.
We obtain exact expressions for the bulk temperature profile
and steady state current in the limit $N \rightarrow \infty$. 
These depend on 
whether $N$ is odd or even. 
  We also study similar temperature oscillations in the NESS of systems with 
noise in the  dynamics.
 These die out as $N \rightarrow \infty$.
 \end{abstract}
\maketitle
\section{Introduction}
The study of non-equilibrium steady states (NESS) of macroscopic systems in contact with heat baths at different temperatures has a long history \cite{dhar08,bonleb, LLP}.  There are no known analytic solutions for interacting Hamiltonian systems, except for harmonic crystals. When the atoms in the crystal all have the same mass, this NESS can be obtained explicitly \cite{RLL,Nak}.  It gives a uniform "temperature", i.e $\langle p_j^2 \rangle /m $, in the bulk of the system.  It also gives heat currents that are independent of the size of the system corresponding to the fact that phonons can travel freely through the crystal. This behavior of the heat current is also true for harmonic systems with periodic arrangement of masses \cite{casher71,Leb-Conn}.  It was therefore 
surprising when numerical simulations of the NESS  showed  
that the bulk temperature profile of a chain with alternating masses oscillates 
between two values and that these 
oscillations did not seem to decay on increasing the system size \cite{dhar08}. 

Here we present analytical solutions for the temperature profile and current in the alternate-mass chain connected to Langevin type heat baths, which prove that the oscillations persist in the $N \to \infty$ limit. Only for very special choice of parameter values, can the oscillations be made to vanish. Surprisingly, the values of the oscillating temperature and of the current depends on whether $N$ is even or odd even in 
the asymptotic system size limit. 
An oscillating temperature profile in a thermodynamically large system 
is surprising when we think from the standard view-point of heat flow occuring
from hot to cold regions.  However this expectation will be true only in  
systems exhibiting local thermal equlibrium where one can define a meaningful 
thermodynamic local temperature. This is the case for a system with normal
diffusive heat transport, though a  microscopic derivation of the conditions
when this is achieved is in general difficult \cite{bonleb}. 
In the study of systems in NESS it is
natural to define a local ``temperature'' from the mean local kinetic energy and
this is what we do here --- the absence of local equilibrium in the 
harmonic chain allows for the ``temperature'' profile to show the
unexpected oscillatory feature.

In the context of testing Fourier's law and investigating the size-dependence of the current, various one-dimensional models with alternate masses have been studied \cite{altmass,Hurt,Dhar-FPU}.
Temperature oscillations have earlier been observed in the steady state of the
alternate mass hard particle gas \cite{Hurt} and in the Fermi-Pasta-Ulam chain
\cite{Dhar-FPU} but in these cases the oscillations decay with system size. 
The case of temperature oscillations persisting for infinite system sizes  is 
thus special to harmonic systems where heat is transmitted by non-interacting
phonons. It is expected that introduction of phonon-phonon interactions will
in general make things different. Here we investigate this issue by considering
alternate-mass harmonic chains where the dynamics is stochasticlly perturbed
by noise which either conserves both energy  and momentum or conserves only
energy. Finally to consider the effect of dimensionality, we present results
from simulations of two-dimensional strips of alternate mass harmonic systems.

The plan of the paper is as follows. In Sec.~(\ref{model}) we define the 
precise model and present some of the numerical results for small finite
systems.  In Sec.~(\ref{exact}) we present the analytic and numerical results 
in the limit $N \to \infty$. In Sec.~(\ref{noise}) we present simulation 
results on temperature profiles in harmonic chains with noisy dynamics. 
In Sec.~(\ref{summary}) we summarize our results and  give a physical
explanation of the results. The
details of our analytical calculations are given in Appendix~(\ref{appa}).  

\section{Model and numerical results for small system sizes}
\label{model}
We consider a one-dimensional chain of $N$ particles labeled $i=1,\ldots,N$ that are placed in an external  harmonic potential (with spring constant $k_o$) and which are interacting with each other through a nearest neighbor harmonic potential (with spring constant $k$). 
Let the vectors $q=(q_1,q_2,\ldots,q_N)$ and $p=(p_1,p_2,\ldots,p_N)$ denote respectively the displacement and momenta of the $N$ particles of the chain. 
The Hamiltonian for the 1D chain we consider is given by: 
\begin{eqnarray}
H &=& \frac{1}{2} \displaystyle\sum_{i=1}^{i=N} p_{i}^{2}/2m_{i} + \frac{1}{2}\displaystyle\sum_{i=1}^{N+1} k (q_{i} - q_{i-1})^2 + \frac{1}{2} \displaystyle \sum_{i=1}^{N} k_o q_{i}^2  \\
&=& \frac{1}{2} p.{\bf M}^{-1}.p + \frac{1}{2} q. {\boldsymbol{\Phi}}.q ~, 
\end{eqnarray}
with $q_0=q_{N+1} =0$, and in the second line we have used a compact notation with ${\bf M}$ defining the mass matrix and ${\boldsymbol \Phi}$ the force-matrix. The ends of the chain are coupled to Langevin reservoirs at temperatures $T_L$ and $T_R$.  The equations of motion of the system is given by:
\bea
m_i \ddot{q}_i= -\sum_{j=1,N} \Phi_{i,j} q_j +\delta_{i,1} 
[-\gamma_L \dot{q}_1  + (2 \gamma_L T_L)^{1/2} \eta_L ]+
\delta_{i,N} [-\gamma_R \dot{q}_N  + (2 \gamma_R T_R)^{1/2} \eta_R ]~,~~\label{langeq}
\eea
for $i=1,2,\ldots,N$, where $\eta_L, \eta_R$ are Gaussian white noises chosen
from distributions with averages  $\la \eta_L(t)\ra=\la \eta_R(t) \ra =0$ and
correlations $\la \eta_L(t) \eta_L(t') \ra = \la \eta_R(t) \eta_R(t') \ra= \delta(t-t')$ and $\gamma_L,\gamma_R$ are dissipation
constants ( 
Note that the derivation in \cite{casher71} uses a different convention for 
the reservoir coupling. The dissipative forces on the end particles were there  taken to be
$-\lambda_1 p_1$ and $-\lambda_N p_N$ and so their coupling constants are
related to ours as $\lambda_1 m_1=\gamma_L, \lambda_N m_N =\gamma_R$ ).
We will be interested in the case where the masses $m_i$
alternate between two values $m_a$ and $m_b$ on successive sites. 

Corresponding to the Langevin equations in Eq.~(\ref{langeq}) it is straightforward 
to write the Fokker-Planck equation to describe the evolution of the phase 
space distribution $\mu(x,t)$, $x=(q_1,\cdots,q_N,p_1,\cdots ,p_N)$. Following
standard methods \cite{risken} it can be shown that the Fokker-Planck equation 
is given by: 
  \begin{equation}
\frac{\partial \mu}{\partial t} + \sum_{i=1}^N \left[ \f{p_i}{m_i} \f{\partial \mu}{\partial q_i}  -\sum_{j=1}^N \Phi_{i,j} q_j \f{\partial \mu}{\partial p_i}\right] = \displaystyle\sum_{i =1 , N} \frac{\gamma_i}{m_{i}} \frac{\partial}{\partial p_{i}} \left [ p_{i} \mu +T_{i}m_{i} \frac{\partial \mu}{\partial p_{i}} \right ]~, 
\label{LV}
\end{equation}
where the right hand side of Eq.~(\ref{LV}) describes the 
interaction of the end particles with the heat baths and $T_{1,N}=T_{L,R}$ and $\gamma_{1,N}=\gamma_{L,R}$.
Let us define the $2N \times 2N $ matrix   
\bea
{\bf a} = \left ( \begin{array}{cc}
\bf{0} & \bf{-M^{-1}} \\
\Phi & \bf{M^{-1}\Gamma} \\
\end{array}  \right )~,
\eea
where ${\bf \Gamma}$ is a $N \times N$ diagonal matrix with $\Gamma_{ij} = \gamma_i \delta_{ij}(\delta_{i1} + \delta_{iN})$ . We also define the $2 N \times 2 N$ matrix ${\bf d}$ with elements $d_{ij} =  2 \gamma \delta_{ij} (T_L \delta_{i,N+1} +T_R \delta_{i,2N})$. 
\begin{figure}
\includegraphics[scale=0.5]{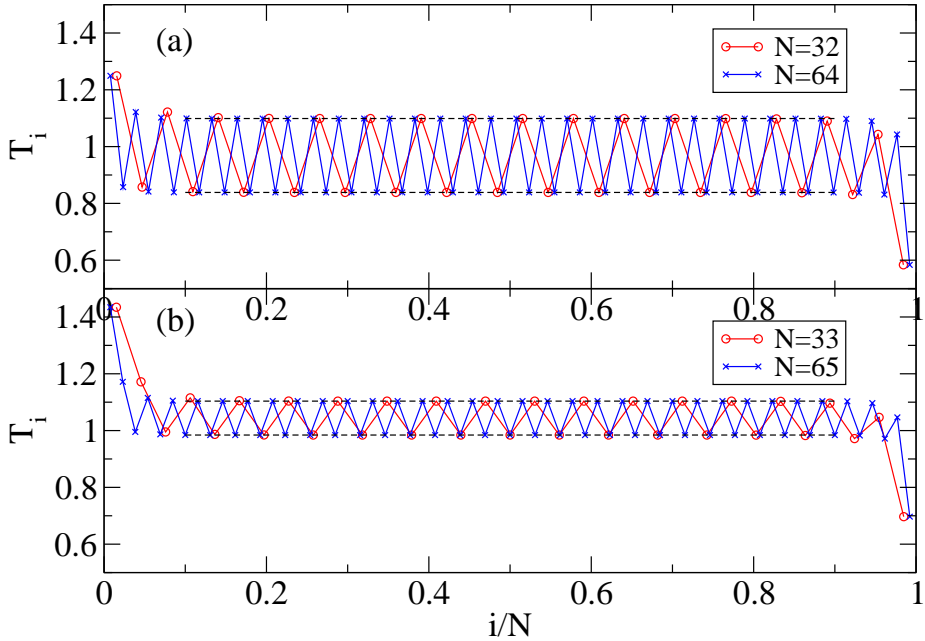}
\caption{(Color online) Temperature profiles for (a) system with even number of sites 
$N=32,64$, and with $\gamma_L=\gamma_R=1.0$ and (b) system with odd number of sites $N=33,65$, and with $\gamma_L=1.5, \gamma_R=0.5$. Other parameters were set to $m_a=0.75, m_b =0.25, k=1, T_L=1.5, T_R=0.5$. The mass of the first particle is always taken to be $m_a$. Note that in (a), the heavier particles are hotter, while in (b), the lighter particles are hotter. The horizontal dashed lines indicate the analytic predictions for $N \to \infty$, from Eqs.~(\ref{int-even},\ref{int-odd}).}
\label{fig1}
\end{figure}
It is known that the steady state distribution is  Gaussian \cite{RLL} and given by
$$ \mu_{s} = (2 \pi)^{-N} Det[{\bf b}]^{-1/2} exp(-\frac{1}{2} \displaystyle\sum_{i,j=1}^{2N} b_{ij}^{-1}x_{i}x_{j})~,$$ 
where the covariance matrix {\bf b} with elements $b_{ij}= <x_i x_j> $ satisfies
\begin{equation}
{\bf a.b +b.a^{\dagger} =d}~. \label{mat-eqn}
\end{equation} 
The solution of the linear equations, Eq.(\ref{mat-eqn}), gives us all the correlations
$b_{ij}$  and hence the temperature profile $T_i =<p_i^2>/m_i
=b_{N+i,N+1}/m_i$ , and the current, $J = k \langle(q_{i+1} -q_i)p_i/m_i \rangle = k (b_{i+1,N+i} - b_{i,N+i})/m_i$.
 In the equal mass case the covariance matrix for $N$ sites can be obtained in a fairly explicit form \cite{RLL}. This seems to be difficult for the alternate mass case. However the matrix equations 
can be solved  numerically  for small system sizes and we can obtain accurate results for the temperature profile and current for these system sizes. 
In Fig.~(\ref{fig1}) we show typical temperature profiles for alternate mass chains with even and odd number of sites for particular choices of parameter
values and $N$. 
We see  oscillations in the temperatures of the particles in the bulk in both 
the even and odd cases, and the amplitude of the oscillations does not seem to
change with system size. 
In the next section, we will obtain expressions for the current and the bulk 
temperatures and show that the temperature ocscillations persist  in the 
$N \rightarrow \infty$ limit.

\section{Analytical and numerical results in $N \to \infty$ limit}
\label{exact}
To obtain analytic results in the limit $N \to \infty$ we follow  \cite{casher71} and  express the covariances in terms of integrals over frequencies. The integrands involve elements of the following Green's function:
\begin{eqnarray}
{\bf G}^+= [ -{\bf M} \omega^2 + {\bf \Phi} - i \omega {\bf \Gamma}]^{-1}~.
\end{eqnarray}
Here we are interested in the temperature and current and these are given by
\cite{casher71,dharroy06} 
\begin{eqnarray}
T_i &=& \frac{1}{\pi} m_i \left[ \gamma_L T_L \displaystyle \int_{-\infty}^{\infty}~ d \omega~ \omega^2~ |{\bf G}^+_{i1}(\omega)|^2  
+  \gamma_R T_R \displaystyle \int_{-\infty}^{\infty}~ d \omega~ \omega^2~ |{\bf G}^{+}_{iN}(\omega)|^2 \right]~,~i=1,2,\ldots,N~~~~\nn \label{temp} \\
J &=& \frac{\gamma_L \gamma_R (T_L-T_R)}{\pi} \int_{-\infty}^\infty d \omega  
~\omega^2~ |{\bf G}^+_{1,N}(\omega)|^2 \label{curr}~.
\end{eqnarray}
We rewrite the above expressions in the following form.
\begin{eqnarray}
T_i &=& I_{i}~T_L+ \hat{I}_{i}~T_R ~,\nn \\
J &=& \frac{\gamma_R}{m_N} (T_L-T_R) I_{N}~, \nn \\
\text{ where the}  \nn \\ ~I_i &=& \frac{m_i \gamma_L}{\pi}  \displaystyle \int_{-\infty}^{\infty}~ d \omega~ \omega^2~ |{\bf G}^+_{i1}(\omega)|^2  ~,~~
\hat{I}_i=\frac{m_i \gamma_R}{\pi}  \displaystyle \int_{-\infty}^{\infty}~ d \omega~ \omega^2~ |{\bf G}^+_{iN}(\omega)|^2~.  
\end{eqnarray}
are independent of the temperatures $T_L$ and $T_R$. Now we note that for the equilibrium case $T_L=T_R$, we must have the same temperature at all sites, {\emph {i.e}} $T_i=T$, and hence deduce the equality 
$I_i+\hat{I}_i=1$.  
Using this fact and defining $T_L=T+\Delta T/2,  T_R=T-\Delta T/2$ we can rewrite the equation for the temperature profile in the following form:
\begin{eqnarray}
T_i=T+(I_i-1/2) \Delta T=T_R +I_i \Delta T.
\end{eqnarray}
We thus only need to evaluate the integral $I_i$, in the limit $N \to \infty$.

So far our treatment has been quite general. We now focus on the alternate mass 
case. We define the first mass to be $m_1=m_a$ and the next to be $m_b$ and so on. Thus odd sites have $m_i=m_a$ and even sites, $m_i =m_b$.  For simplicity we only consider the unpinned case $k_o=0$. It is straightforward to extend the calculations to the case $k_o\neq0$. 
Without loss of generality we can choose time and energy scales so that $k=1$
and $m_a+m_b=1$. 
We give the details of the calculation in the appendix. The main result is that $I_i$ can be 
written as a sum of two parts, one coming from the acoustic modes of the system
and one from the optical modes.  We note that the mode frequencies for the acoustic and optical bands are respectively given by:
$\omega_-^2 = ({1}/{m_am_b}) [1-\phi(q)]~, \omega_+^2 = ({1}/{m_a m_b}) [1+\phi(q)]$, where $\phi(q) = [1-2m_a m_b(1-\cos q)]^{1/2}~,~,0\leq q \leq \pi$.
The various expressions depend on whether $N$ is even or odd and for these two cases corresponding to superscript $E,O$ respectively, we get:

{\bf Case (1)}- $N=2L$, $L \rightarrow \infty$:
\begin{eqnarray}
T^{E}_{o} &=& T + \Delta T \left[  \int_{0}^{\pi} dq \f{\gamma_L m_a}{2 \pi \phi(q)}  \frac{ (m_b\omega_{+}^2-2)^2 + 
4 \gamma_R^2 \omega_{+}^2   \text{cos}^2 (q/2)} { |2(\gamma_L+\gamma_R) - (m_a\gamma_R+m_b\gamma_L) \omega_+^2| ~(1+ \gamma_L \gamma_R \omega_{+}^2 ) }  \nn \right. \\ &&
\left.     
+ \int_{0}^{\pi} dq \f{\gamma_L m_a}{2 \pi \phi(q)}  \frac{ (m_b\omega_{-}^2-2)^2 + 
4 \gamma_R^2 \omega_{-}^2   \text{cos}^2 (q/2)} { |2(\gamma_L+\gamma_R) - (m_a\gamma_R+m_b\gamma_L) \omega_-^2| ~(1+ \gamma_L \gamma_R \omega_{-}^2 )}
 -\f{1}{2} \right]~, \nn \\
T^{E}_{e} &=& T+  
\Delta T \left[ \int_{0}^{\pi} dq \f{\gamma_L m_b}{2\pi \phi(q)} \f{ 4 \text{cos}^2 (q/2)  + \gamma_R^2 \omega_{+}^2 (m_a \omega_{+}^2-2)^2 } {|2(\gamma_L+\gamma_R) -(m_a\gamma_R+m_b\gamma_L)\omega_+^2|~ (1+ \gamma_L \gamma_R \omega_{+}^2) } \nn \right. \\ && \left. + 
\int_{0}^{\pi} dq \f{\gamma_L m_b}{2\pi \phi(q)} \f{ 4 \text{cos}^2 (q/2)  + \gamma_R^2 \omega_{-}^2 (m_a \omega_{-}^2-2)^2 } {|2(\gamma_L+\gamma_R) -(m_a\gamma_R+m_b\gamma_L)\omega_-^2|~ (1+ \gamma_L \gamma_R \omega_{-}^2) }
 -\f{1}{2} \right]~, \nn \\
  J^{E}   &=& \Delta T  \left[ \int_{0}^{\pi} dq \f{\gamma_L \gamma_R}{\pi \phi(q)} \f{\text{sin}^2 q}{ |2(\gamma_L+\gamma_R) - (m_a\gamma_R+m_b\gamma_L) \omega_{+}^2|~ (1+\gamma_L \gamma_R \omega_{+}^2) }  \right. \nn \\ && \left. 
+ \int_{0}^{\pi} dq \f{\gamma_L \gamma_R}{\pi \phi(q)} \f{\text{sin}^2 q}{ |2(\gamma_L+\gamma_R) - (m_a\gamma_R+m_b\gamma_L) \omega_{-}^2|~ (1+\gamma_L \gamma_R \omega_{-}^2) }
\right]~, \label{int-even}
  \end{eqnarray}   
where the subscript $o$ refers to odd sites and $e$ to even sites.

{\bf Case (2)}-  $N=2L+1$, $L \rightarrow \infty$.
\begin{eqnarray}
T^{O}_{o} &=& T+ \Delta T \left[
 \int_{0}^{\pi} dq \f{\gamma_L m_a}{2(\gamma_L+\gamma_R) \pi \phi(q)}~ \f{4 \text{cos}^2 (q/2) +  \gamma_R^2 \omega_{+}^2 (m_b \omega_{+}^2 -2)^2}{ | (m_a \omega_{+}^2 -2) + \gamma_L \gamma_R \omega_{+}^2 (m_b \omega_{+}^2 -2)|} \right. \nn \\ 
&& \left. + \int_{0}^{\pi} dq \f{\gamma_L m_a}{2(\gamma_L+\gamma_R) \pi \phi(q)}~ \f{4 \text{cos}^2 (q/2) +  \gamma_R^2 \omega_{-}^2 (m_b \omega_{-}^2 -2)^2}{ | (m_a \omega_{-}^2 -2) + \gamma_L \gamma_R \omega_{-}^2 (m_b \omega_{-}^2 -2)|}
 - \f{1}{2} \right]~, \nn \\
T^O_{e} &=& T+\Delta T  \left[ \int_{0}^{\pi} dq \f{\gamma_L m_b}{2 (\gamma_L+\gamma_R)  \pi \phi(q)}~ \f{ (m_a \omega_{+}^2 -2)^2 +  4 \gamma_R^2 \omega_{+}^2 {cos}^2 (q/2)} { | (m_a \omega_{+}^2 -2) + \gamma_L \gamma_R \omega_{+}^2 (m_b \omega_{+}^2 -2)|} \right. \nn \\
&& \left. + 
\int_{0}^{\pi} dq \f{\gamma_L m_b}{2 (\gamma_L+\gamma_R)  \pi \phi(q)}~ \f{ (m_a \omega_{-}^2 -2)^2 +  4 \gamma_R^2 \omega_{-}^2 {cos}^2 (q/2)} { | (m_a \omega_{-}^2 -2) + \gamma_L \gamma_R \omega_{-}^2 (m_b \omega_{-}^2 -2)|}
-\f{1}{2} \right]~, \nn \\ 
J^{O} &=& \Delta T  \left[ \int_{0}^{\pi} dq \f{\gamma_L \gamma_R}{(\gamma_L+\gamma_R) \pi \phi(q)} ~ \f{\text sin^2 q}{|(m_a \omega_{+}^2 -2) + \gamma_L \gamma_R \omega_{+}^2  (m_b \omega_{+}^2 -2) |} 
\right. \nn \\
&& + \left. \int_{0}^{\pi} dq \f{\gamma_L \gamma_R}{(\gamma_L+\gamma_R) \pi \phi(q)} ~ \f{\text sin^2 q}{|(m_a \omega_{-}^2 -2) + \gamma_L \gamma_R \omega_{-}^2  (m_b \omega_{-}^2 -2) |} 
 \right]~. \label{int-odd}
\end{eqnarray}

\begin{figure}
\includegraphics[scale=0.5]{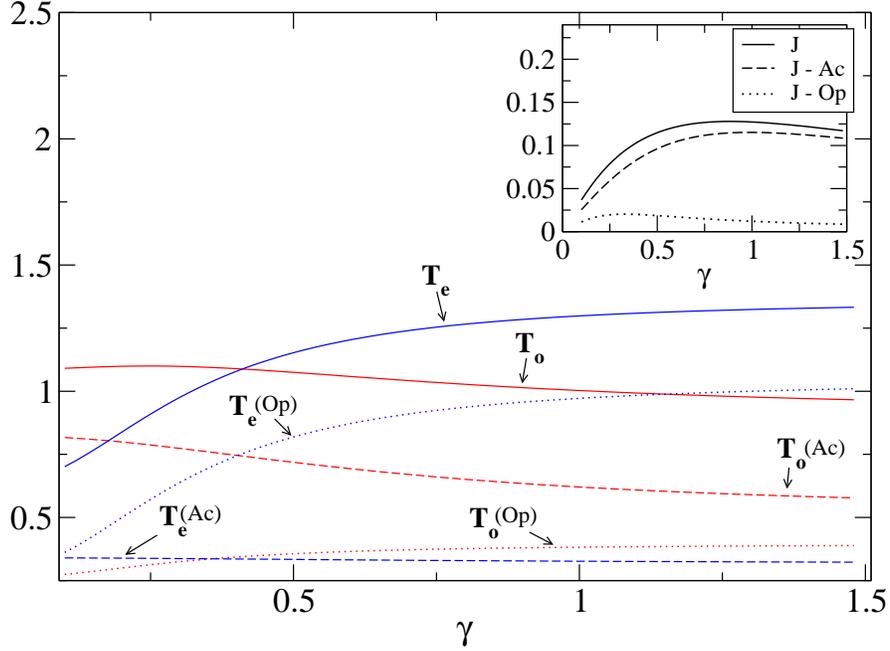}
\caption{(Color online) Temperatures at odd and even sites for a chain with even number of particles, plotted as a function of $\gamma=\gamma_L=\gamma_R$, $m_a=0.75, m_b =0.25, k=1, T_L=1.5, T_R=0.5$.  
We also plot separately the contributions of the acoustic and optical modes to 
the temperature at any site. The inset shows $J$ and also the contributions of the acoustic and optical modes. 
}
\label{fig2}
\end{figure}   
\begin{figure}
\includegraphics[scale=0.5]{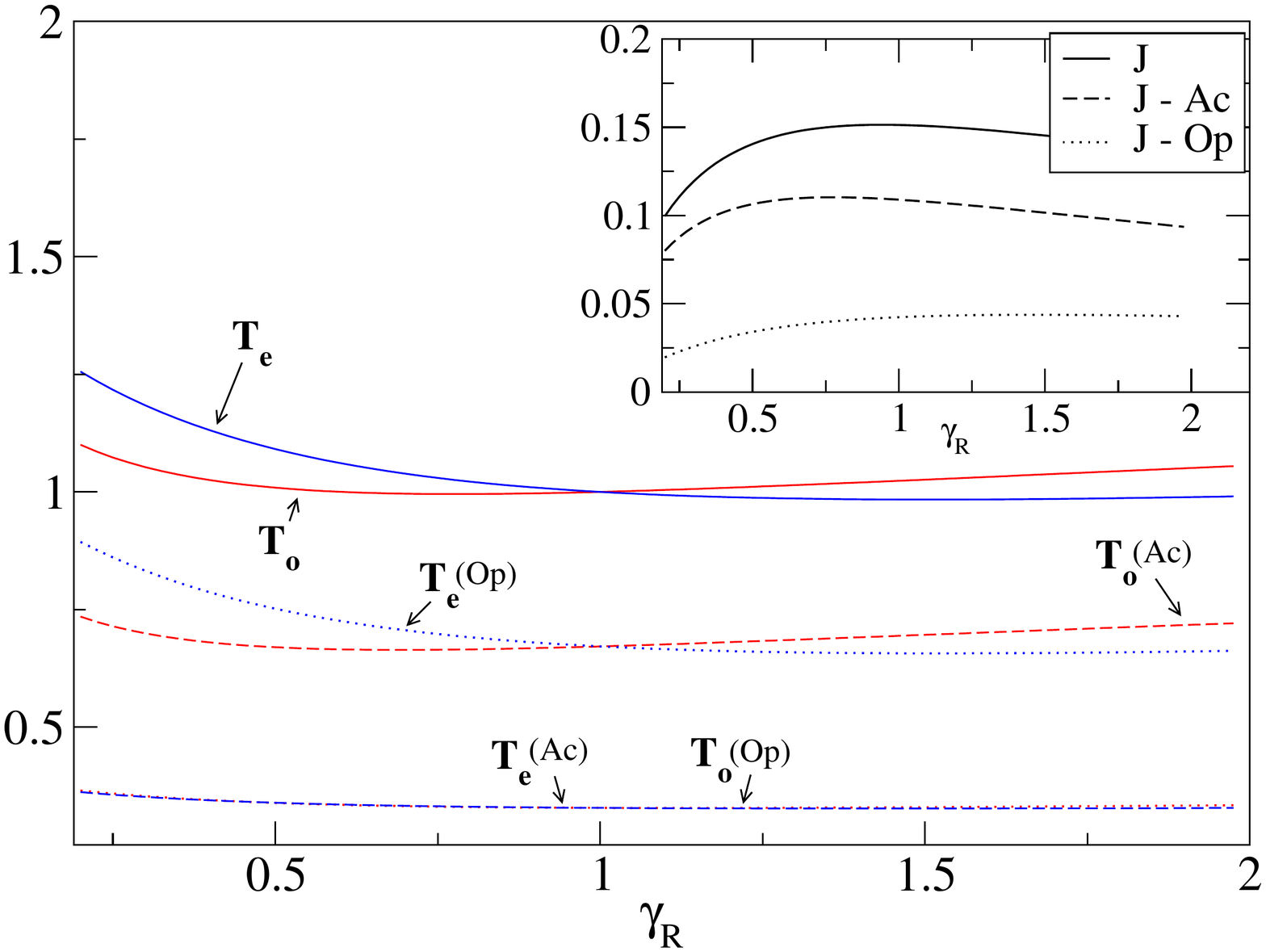}
\caption{(Color online) Temperatures at odd and even sites for a chain with odd number of particles, plotted as a function of $\gamma_R$ with $\gamma_L=1$, $m_a=0.75, m_b =0.25, k=1, T_L=1.5, T_R=0.5$.  
We also plot separately the contributions of the acoustic and optical modes to 
the temperature at any site. The inset shows the variation of heat current with $\gamma_R$ and also the contributions of the acoustic and optical modes. 
  }
\label{fig3}
\end{figure}   

We now present some numerical data for the two cases of even $N$ and odd $N$
for various parameter sets. When $\gamma_L=\gamma_R$, the above integrals can
be carried out exactly, see Eqs.~(\ref{tempo-even}-\ref{curr-odd}). In other
cases, we evaluated the integrals numerically (using Mathematica). 

{\bf Case (1)}:  We consider chains with even 
$N$ and set $\gamma_L=\gamma_R=\gamma$. In Fig.~(\ref{fig2}) we plot the temperatures on the odd ($T^E_o$) and even ($T^E_e$) sites, and also the current ($J^E$ in inset) as a function 
of the parameter $\gamma$. We also separately plot the contributions of the 
acoustic and optical modes to the temperatures and current.
We note the following features:

(i) Depending on the value of $\gamma$, either the heavier particles 
(those on odd sites),  or the lighter ones are hotter. 
At $\gamma \approx 0.41$, the temperatures at the odd and even sites are equal. 

(ii) The temperature of the heavier particles gets its main contribution from 
the acoustic modes while that of the lighter particles comes mostly from the 
optical modes. The heat current is mostly carried by the acoustic modes.  

{\bf Case (2)}:  We consider chains with odd  
$N$. In this case, $\gamma_L=\gamma_R$ becomes a very special case: the masses 
of the end particles being equal, this condition implies  
symmetry between the left and right reservoirs, and this leads to  a 
uniform bulk temperature equal to $(T_L+T_R)/2$. The more typical 
situation is when the two couplings are different and we consider this 
by setting $\gamma_L=1$ and changing $\gamma_R$. 
In Fig.~(\ref{fig3}) we plot the temperatures on the odd ($T^O_o$) and even ($T^O_e$) sites, and also the current ($J^O$ in inset) as a function 
of the parameter $\gamma_R$. We also separately plot the contributions of the 
acoustic and optical modes to the temperatures and current.
We note the following features:

(i) Depending on the value of $\gamma_R$, either the heavier particles 
(those on odd sites),  or the lighter ones are hotter. 
At a special value of $\gamma_R =\gamma_L$, the temperatures at the odd and even sites are the same. They are both equal to the mean temperature $T=1$.

(ii) As for the even $N$ case, here also we see that the temperature of the 
heavier particles gets it main contribution from 
the acoustic modes while that of the lighter particles comes mostly from the 
optical modes. The heat current is again mostly carried by the acoustic modes.

\section{Simulation results on the effect of noise in the dynamics}
\label{noise}

\begin{figure}
\includegraphics[scale=0.4]{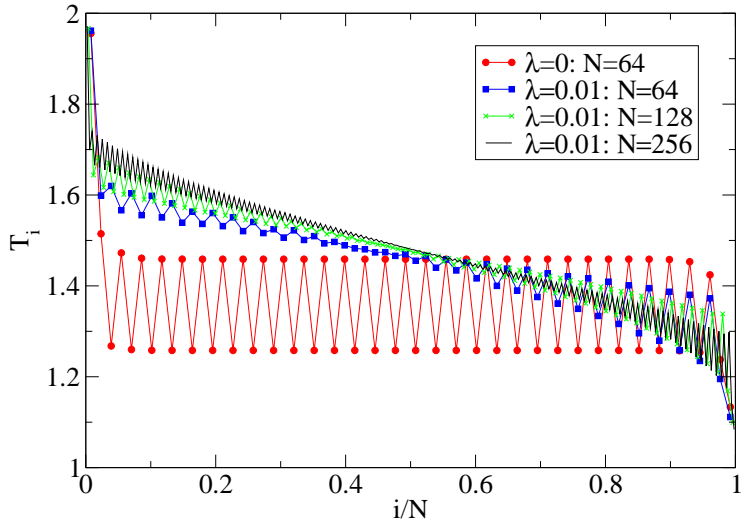}
\caption{(Color online) Temperature profiles with energy-- and momentum--conserving noisy dynamics for a harmonic chain with even number of particles. Other parameters were taken to be 
$m_a=0.5$, $m_b=1.5$, $\gamma_L=\gamma_R=1.0$ and $T_L=2.0, T_R=1.0$.}
\label{fig:EN}
\end{figure}
\begin{figure}
\includegraphics[scale=0.4]{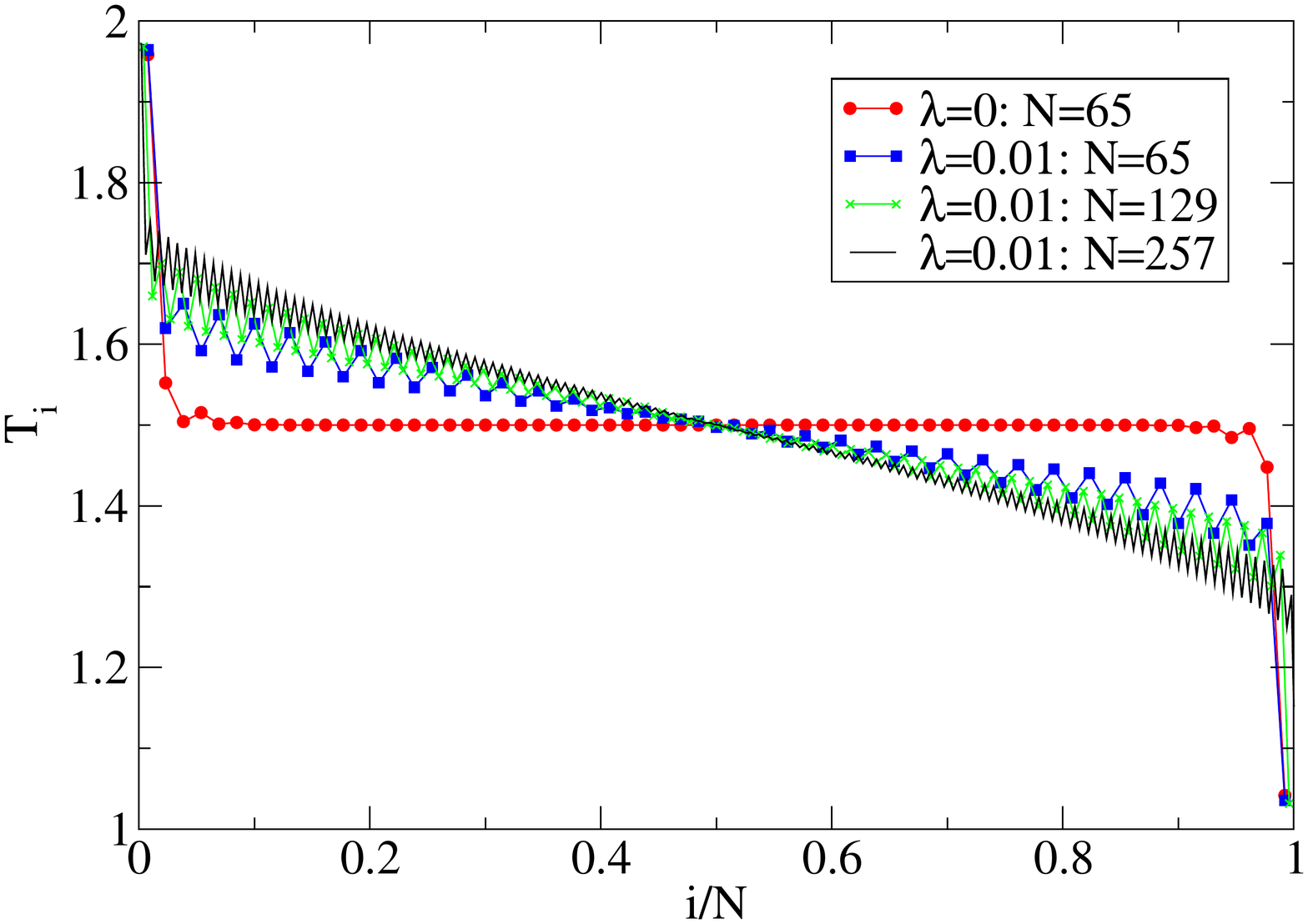}
\caption{Temperature profiles with energy-- and momentum--conserving noisy dynamics for a harmonic chain with an odd number of particles. Other parameters were taken to be $m_a=0.5$, $m_b=1.5$, $\gamma_L=\gamma_R=1.0$ and $T_L=2.0, T_R=1.0$.}
\label{fig:ON}
\end{figure}

\begin{figure}
\includegraphics[scale=0.4]{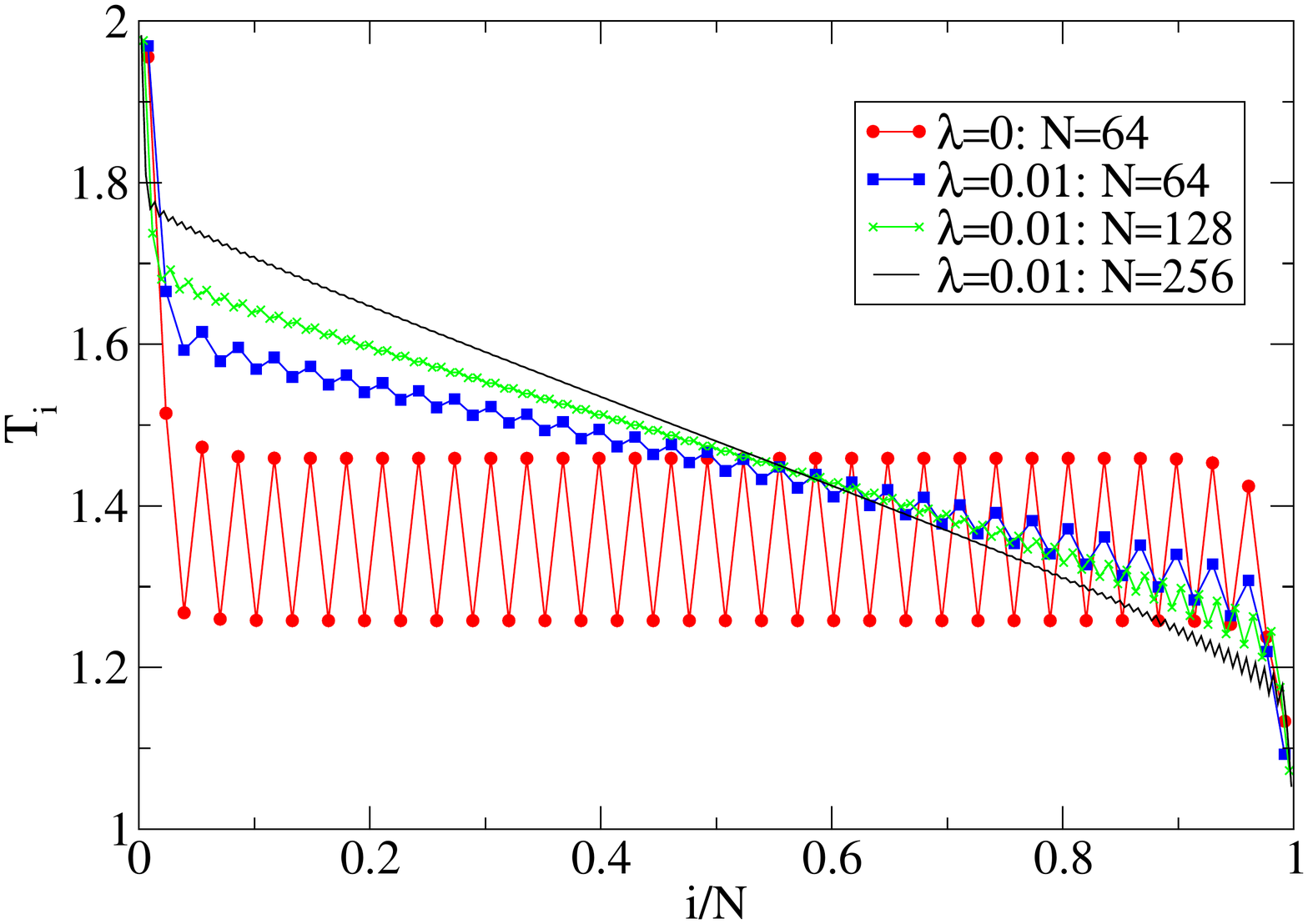}
\caption{(Color online) Temperature profiles with only { energy--conserving noisy dynamics} for a harmonic chain with even number of particles. Other parameters were taken to be
$m_a=0.5$, $m_b=1.5$, $\gamma_L=\gamma_R=1.0$ and $T_L=2.0, T_R=1.0$.}
\label{fig:ENN}
\end{figure}
\begin{figure}
\includegraphics[scale=0.4]{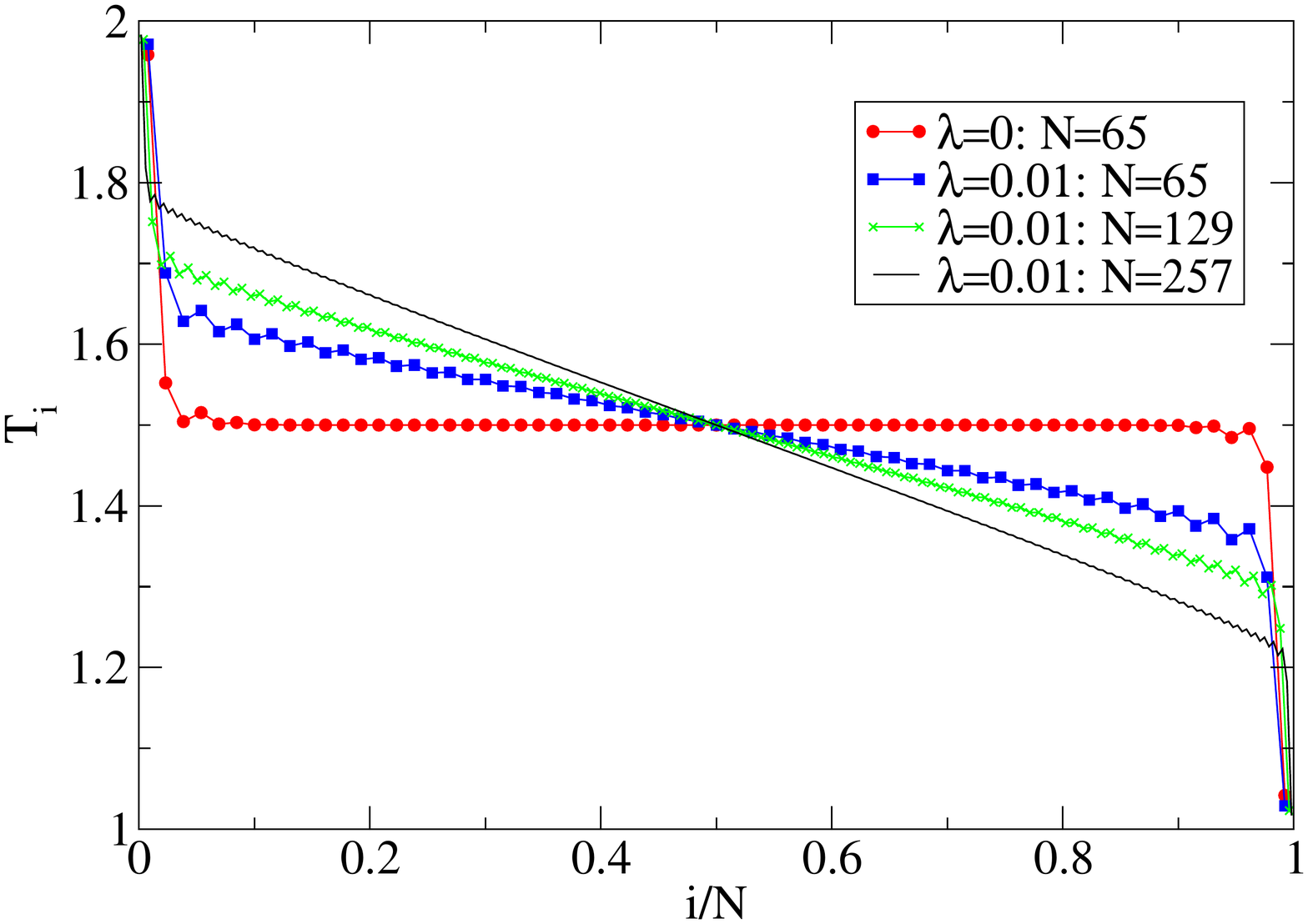}
\caption{(Color online) Temperature profiles with only {energy--conserving noisy dynamics} for a harmonic chain with an odd number of particles. Other parameters were taken to be $m_a=0.5$, $m_b=1.5$, $\gamma_L=\gamma_R=1.0$ and $T_L=2.0, T_R=1.0$.}
\label{fig:ONN}
\end{figure}
As discussed in the introduction, temperature oscillations have been observed
in anharmonic chains, where however the oscillations decay with system size.  
This is expected since anharmoncity leads to interactions between phonons
which helps to establish local thermal equilibrium. A simple  model 
which incorporates phonon-phonon interactions  was  introduced in
\cite{Bern,Basile} where the determinsitic dynamics of the Harmonic chain is
stochastically perturbed. 
Here we have  carried out simulations with this noisy dynamics and looked at
it's effect on the temperature profiles of the alternating mass chain. 
There are two cases to consider:  

(a) {\emph {Momentum conserving noise}}:  Here, in addition to the 
Hamiltonian dynamics without pinning, one introduces random exchange of momentum between nearest neighbor particles, which occurs with a rate $\lambda$. This conserves both 
momentum and energy. 
In Figs.~(\ref{fig:EN},\ref{fig:ON}) we show the effect of momentum conserving noise on the temperature profiles for chains of even and odd number of particles.
In the even $N$ case we see that, on introducing noise, the size of the 
oscillations has decreased and the phase of the oscillation on the left half has changed sign. For the odd case, the  
choice of parameters ($\gamma_L=\gamma_R=1$) corresponds to a case with no
oscillations when $\lambda= 0$. On introducing noise, $\lambda > 0$, one gets
oscillations very similar to the even $N$ case. We also see that  the
oscillation amplitude  becomes smaller on increasing system size, for both
even and odd $N$ cases. 

Thus we see that the temperature profile in this system with energy-momentum conserving noisy dynamics shows the following qualitative features :
 (i) The oscillations decay as we go into the bulk, 
 (ii) There is a phase shift in the sign of the oscillation amplitude as
one crosses the center of the chain. The lighter particles at the hot end 
are always hotter than the heavier particles. At the cold end, the heavier 
particles are hotter. Thus this is {\it qualitatively} different from the 
harmonic case, 
(iii) For large $N$, the temperature profile is not sensitive to whether 
$N$ is even or odd.
These same features have also been observed earlier for 
the alternate mass FPU chain \cite{Dhar-FPU} which had a quartic 
interparticle interaction potential (in addition to the harmonic one).  
The FPU system has momentum conservation and does not satisfy Fourier's Law as
is also the case for the system with noisy dynamics.

(b) {\emph {Momentum non-conserving case}}: 
When the noise only conserves energy but not momentum, as can be obtained by randomly reversing the  velocity of the $i^{th}$ particle at rate $\lambda$, then, as is seen in \cite{LAV}, the NESS for $N \to \infty$ corresponds to a local equilibrium state. This ensures that the $T_i$ in the bulk is the same for $i$ odd or even independently of whether $N$ is even or odd.  In  Figs.~(\ref{fig:ENN},\ref{fig:ONN}), we show the effect of addition of velocity flipping dynamics 
on the temperature profile for harmonic chains with odd and even number of particles. We observe that for the even case, the oscillations in the temperature decreases considerably on introducing the noise, and this reduction is greater when $N$ is larger. For the odd case however, for small systems, introduction of noise produces small oscillations in the temperature profile, but these oscillations eventually decrease as the system-size is increased. For both even and odd total number of particles, the decay of the oscillation amplitude with system size is faster than for the momentum-conserving case and we quickly get a linear temperature profile in the bulk of the system. 

\begin{figure}
\includegraphics[scale=0.4]{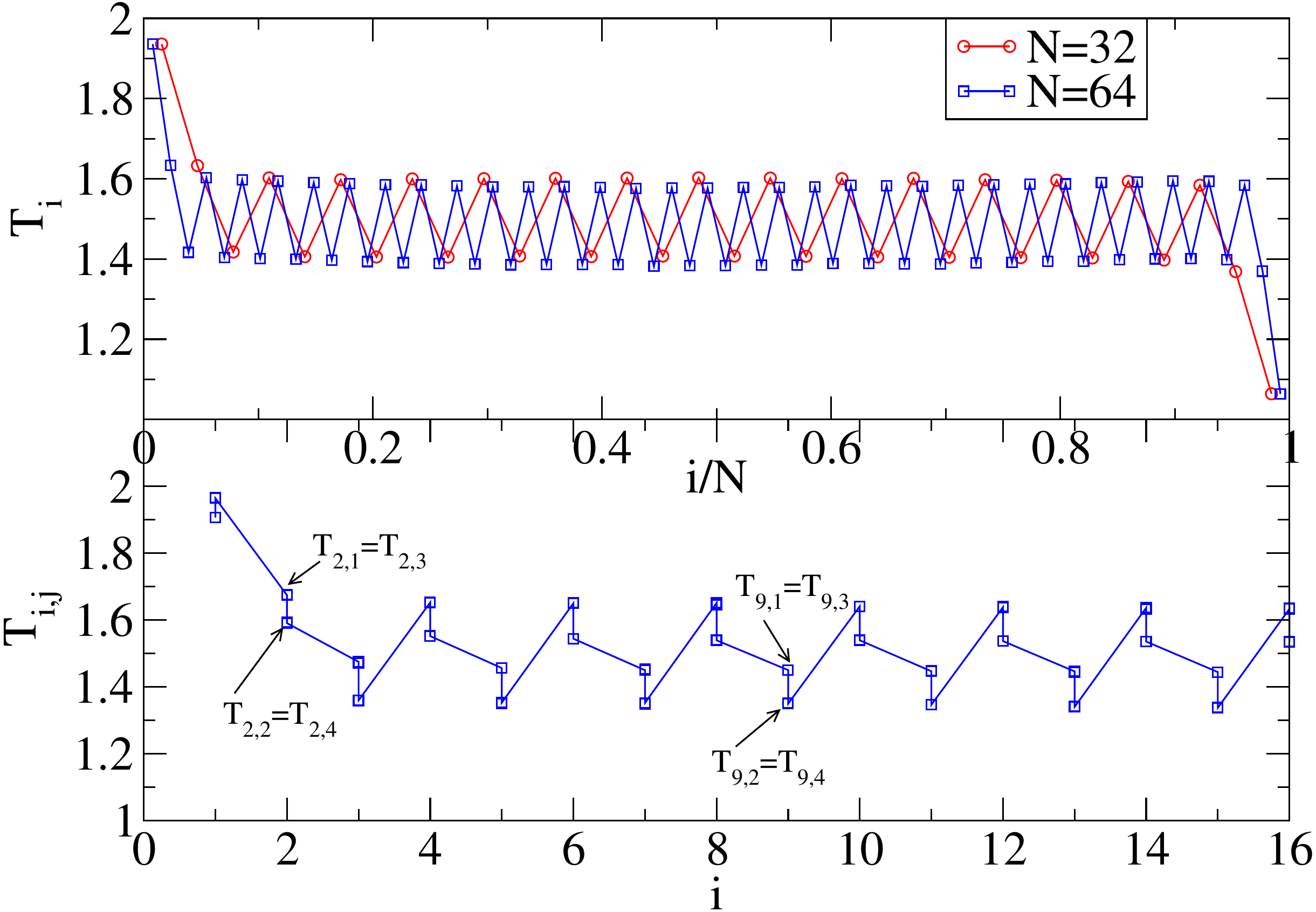}
\caption{(Color online) Simulation results for temperature profile
for a two-dimensional $N\times W$
strip of harmonically coupled particles with a periodic arrangement of
masses.
The sites on the strip are labeled $(i,j)$
with $i=1,\dots,N$ and $j=1,\ldots,W$. Particles at sites with even $i+j$
have
mass $m_a$ and others have masses $m_b$.
Heat baths are attached to all sites
on the layers $i=1$ (temperature $T_L$) and $i=N$ (temperature
$T_R$). Periodic boundary conditions are imposed in the transverse ($j$)
direction.
Upper plot shows the average temperature on succesive layers for chains of
lengths $N=32$ and $N=64$.    There are oscillations in the transverse
direction also and this is shown in the lower plot which shows the
temperatures
$T_{i,j}$ on all sites of a section of the $N=64$ chain.
Note that from symmetry we
have $T_{i,j}=T_{i,j+2}$, and this can be observed here.
The width of the strips were taken to be $W=4$.
The other parameters were taken to be $m_a=0.6$, $m_b=1.4$,
$\gamma_L=\gamma_R=1.0$ and $T_L=2.0, T_R=1.0$.} 
\label{fig:strip}
\end{figure}

\section{Discussion}
\label{summary}

In this paper we  obtained exact expressions for the temperature-profile and the heat current in the alternate mass chain connected to heat baths at different temperatures in the limit of infinite system size. This proves rigorously that the temperature oscillations of successive particles in the bulk persist even  in the thermodynamic limit.  

We provided an understanding of these oscillations by noting that in any given normal mode, the mean kinetic energy of a particle depends on its mass.  
In an  acoustic mode, the heavier particles have higher mean kinetic energy than 
the lighter ones, while in an optical mode, the lighter particles have higher
kinetic energy. 
On connecting the chain to heat reservoirs each of the  modes are 
excited to different degrees, depending on the parameters. 
The kinetic energy of a particle gets contributions from all the modes,
both acoustic and optical and the net result depends on the distribution of energy in the different modes.
If both the baths have the same temperature, we have an equilibrium steady state in which each mode has the same average energy (equipartition). In this case the temperature at all sites are equal.  The same is true locally when the system is in local equilibrium. 

 The situation is different in the non-equilibrium case where we do not have local equilibrium and there is no equipartition of energy between the different modes. We then expect generically that 
the mean kinetic energy (temperatures) obtained by adding the contributions 
of all modes will depend on the mass of the particle.  It is therefore not so
surprising that we get different kinetic energies for the different masses.
From the above explanation we expect that temperature oscillations 
should also occur in higher dimensional periodic harmonic systems.  
Simulation results for two-dimensional strips (see Fig.~(\ref{fig:strip}) 
suggest that this is the case, but more extensive studies are necessary to 
establish the role of dimensionality.

As already noted there will be no oscillations in the bulk if the NESS is in local
thermal equilibrium. To achieve this one introduces interactions between the
phonons. Interactions between phonons can be introduced for example by adding
stochasticity in the dynamics and we studied this case numerically. We find
that in this case the oscillations are  qualitatively different from the
purely harmonic case and do not survive in the  limit of large system size. 

{\it Acknowledgements}: AD thanks DST for support through the Swarnajayanti 
fellowship. The work of VK and JLL was supported in part by NSF Grant  DMR 08-02120 and AFOSR grant AF-FA09550-07. The authors also thank the Fields Institute,
Toronto, where a part of the work was carried out.

\appendix
\section{Details of calculation}
\label{appa}
Here we give more details of the derivation for the temperature profile and 
the current. We basically need to evaluate the integral
\begin{equation}
I_i = \frac{m_i \gamma_L}{\pi}   \int_{-\infty}^{\infty}~ d \omega~ \omega^2~ |{\bf G}^+_{i1}(\omega)|^2 ~,
\end{equation}
where ${\bf G}^+= [-{\bf M} \omega^2 + {\bf \Phi} - \imath \omega {\bf \Gamma}]^{-1}$.
We consider the case with $k=1$.  
Let us   define $\Delta_{l,m}$ as the
determinant of the sub-matrix of $[-{\bf M} \omega^2 +{\bf \Phi}-i \omega{\bf \Gamma}]$ that
starts from the $l^{\rm th}$ row and column and ends in the $m^{\rm
  th}$ row and column. We also define $D_{l,m}$ as the determinant of the
sub-matrix of $[-M \Omega^2 +\Phi]$ starting from the $l^{\rm th}$ row
and column and ending in the $m^{\rm   th}$ row and column. In terms
of these one has:
\bea
{\bf G}^+_{l,1}(\Omega) &=& \frac{\Delta_{l+1,N}}{\Delta_{1,N}},~~{\bf G}^+_{l,N}(\omega)=
\frac{\Delta_{1,l-1}}{\Delta_{1,N}} \label{gsol} \\
{\rm with}~~\Delta_{1,l-1} &=&D_{1,l-1}-i \omega \gamma_L D_{2,l-1} \nn \\
\Delta_{l+1,N}&=&D_{l+1,N}-i \omega \gamma_R D_{l+1,N-1} \nn \\
\Delta_{1,N}&=&D_{1,N}-\imath \omega  (\gamma_R D_{1,N-1}+ \gamma_L D_{2,N})- \omega^2 \gamma_L \gamma_R D_{2,N-1}~.\label{delteq} 
\eea
Let us now define $f(l)=D_{1,2l}$ and $g(l)=D_{1,2l-1}$. 
These satisfy the recursion relation,
\begin{eqnarray}
 \left[ \begin{array}{c} f(l) \\ g(l) \end{array} \right] &=& {\bf B} 
 \left[ \begin{array}{c} f(l-1) \\ g(l-1) \label{rec1} \end{array}   \right]~, \nn \\
{\rm where}~
{\bf B} &=& \left[ \begin{array}{cc}  
(2-m_a\omega^2)(2-m_b \omega^2) - 1 &  -(2-m_b \omega^2) \\                      
  (2-m_a \omega^2)  & -1 \\
 \end{array} \right] \nn ~,
   \end{eqnarray}                                                           
 with the initial condition $f(0)=1$ and $g(0)=0$. Hence we get
\begin{eqnarray}
\left[ \begin{array}{c}  f(l) \\ g(l) \end{array} \right] =  {\bf B}^l  \left[ \begin{array}{c} 1\\ 0  \end{array} \right] ~.
\end{eqnarray} 
The matrix ${\bf B}$ has unit determinant and can be expressed in terms of the
Pauli spin matrices $\vec{\bf \sigma}$ as follows:
\begin{eqnarray}
{\bf B} &=& \cos q ~\textbf{1} + \imath \vec{\boldsymbol \sigma}. \vec{n} \sin q =e^{\imath\vec{\boldsymbol \sigma}. \vec{n} q} \nn \\ 
{\rm where} ~\cos q &=& \text{Tr} \f{{\bf B}}{2} = \f{(2 -m_a \omega^2)(2-m_b\omega^2) -2}{2}~,  \label{om-q}
\end{eqnarray}
and $\vec{n}$ is a three dimensional unit vector. Hence we get:
\begin{eqnarray} 
\textbf{B}^l &=&  =e^{\imath\vec{\boldsymbol \sigma}. \vec{n} ~lq} =  \cos(l q)~\textbf{1} + \sin (lq) \f{{\bf B} - \cos q ~\textbf{1}}{\sin q}  \label{mat-exp2}
\end{eqnarray}      
Combining Eq.~(\ref{mat-exp2}) and Eq.~(\ref{rec1}) we have : \\
\begin{eqnarray}
\label{f-g}
f(l)&=& \frac{\sin (l+1/2)q}{\sin (q/2)}  \label{evendet} \\
g(l)&=& \frac{\sin (l q)}{\sin q} (2-m_a \omega^2) \label{oddD}
\end{eqnarray}       
Note that for odd-dimensional matrices with the first mass equal to $m_b$, the determinant 
would be given by Eq.~(\ref{oddD}) with $m_a$ replaced by $m_b$.
Using these expressions in Eqs.~(\ref{gsol},\ref{delteq}), we then get the following forms for the integrals $I_i$, depending on whether $N$ is even or odd.

  {\bf Case(1)} - even $N$:
\begin{eqnarray}
  I_{odd~i} &=& \f{2 m_a \gamma_L}{\pi} \int_{0}^{\infty} d \omega \omega^2 \frac{\left( \frac{\sin^2 \f{(N-i+1)q}{2}}{\sin^2 q} 
  (2-m_b\omega^2)^2 
     + \gamma_R^2 \omega^2   \frac{\sin^2 \f{(N-i)q}{2}}{\sin^2 (q/2)}  \right)}{|\Delta_{1,N}|^2}~, \nn \\
I_{even~i} &=& \f{2 m_l \gamma_L}{\pi}  \int_{0}^{\infty} d \omega \omega^2 \f{\f{\sin^2 \f{(N-i+1)q}{2}}{\sin^2 (q/2)}  + \gamma_R^2 \omega^2 \f{\sin^2 \f{(N-i)q}{2}}{\sin^2 q} (2 -m_a \omega^2)^2}{|\Delta_{1,N}|^2}~, \nn \\
I_N &=&  \f{2 \gamma_L m_a}{\pi} \int_{0}^{\infty} d \omega 
 \f{\omega^2}{|\Delta_{1,N}|^2}~, \label{IE}  \\
{\rm where} \nn \\
\Delta_{1,N} &=& \left[  \f{\sin \f{(N+1)q}{2}}{\sin (q/2)} - \gamma_L \gamma_R \omega^2 \f{\sin \f{(N-1)q}{2}} {\sin (q/2)} \right] +  \imath \omega \Big[ \gamma_L (2-m_b \omega^2) +\gamma_R ( 2-m_a\omega^2) \Big]  \f{\sin (Nq/2)} {\sin q}~.  \nn
  \end{eqnarray}

  {\bf Case(2)} - odd $N$: 
\begin{eqnarray}
  I_{odd~i} &=& \f{2 m_a \gamma_L}{\pi} \int_{0}^{\infty} d \omega \omega^2 \frac{\left( \frac{\sin^2 \f{(N-i+1)q}{2}}{\sin^2 (q/2)} 
     + \gamma_R^2 \omega^2   (2-m_b\omega^2)^2   \frac{\sin^2 \f{(N-i)q}{2}}{\sin^2 (q)}  \right)}{|\Delta_{1,N}|^2}~, \nn \\
I_{even~i} &=& \f{2 m_l \gamma_L}{\pi}  \int_{0}^{\infty} d \omega \omega^2 \f{(2 -m_a \omega^2)^2 \f{\sin^2 \f{(N-i+1)q}{2}}{\sin^2 q)}  + \gamma_R^2 \omega^2 \f{\sin^2 \f{(N-i)q}{2}}{\sin^2 (q/2)}} 
{|\Delta_{1,N}|^2}~, \nn \\
I_N &=&  \f{2 \gamma_L m_a }{\pi} \int_{0}^{\infty} d \omega 
 \f{\omega^2}{|\Delta_{1,N}|^2}~,   \label{IO} \\
{\rm where}   \nn  \\
\Delta_{1,N} &=& \left[  (2-m_a \omega^2) \f{\sin \f{(N+1)q}{2}}{\sin q} - \gamma_L \gamma_R \omega^2 (2-m_b \omega^2) \f{\sin \f{(N-1)q}{2}} {\sin q} \right] +  \imath \omega \Big( \gamma_L + \gamma_R \Big)  \f{\sin (Nq/2)} {\sin q}~. \nn
  \end{eqnarray}

We now consider points in the bulk such that $x=i/N$ and $(N-i)/N$ remain 
finite in  the $N \to \infty$ limit. 
We now note that, for real values of $ 0< q < \pi$, Eq.~(\ref{om-q}) has two 
allowed solutions for $\om$, namely:
\begin{eqnarray}
\omega_-^2 &=& \f{1}{m_am_b} [1-\phi(q)] \nn \\
\omega_+^2 &=& \f{1}{m_a m_b} [1+\phi(q)]~, \nn \\
{\rm where}~\phi(q) &=& [1-2m_a m_b(1-\cos q)]^{1/2}~,~,0\leq q \leq \pi~. \nn
\end{eqnarray}
 These correspond to the frequencies in the acoustic and optical branches of the
lattice with the frequency ranges $0<\om_- < \sqrt{2/M}$ and $\sqrt{2/m} < \om_+ < \sqrt{2/(mM)}$, where $m$ ($M$) is the smaller (larger) of the two masses. For frequencies outside these ranges, Eq.~(\ref{om-q}) gives imaginary 
values of $q$. This means that, for these frequencies, terms such as 
$\sin{Nx q}$ grow exponentially with $N$. Hence it is clear that, in the limit
$N \to \infty$, the integrals in Eqs.~(\ref{IE},\ref{IO}) only get contributionsfrom frequencies in the acoustic and optical bands.  Thus for each of the integrals above, we get:
\bea
\int_0^\infty d\omega F(\omega) &=& \int_0^{\sqrt{2/M}} d\om_-~ F(\omega_-) 
+\int_{\sqrt{2/m}}^{\sqrt{2/(mM)}} d\omega_+ ~F(\omega_+)  \nn \\
&=& \int_0^\pi dq \big| \f{d\om_-}{dq} \big| -~ F(\omega_-(q)) 
+\int_0^\pi dq \big| \f{d\omega_+}{dq} \big| ~F(\omega_+(q))  \nn 
\eea 

We now note from Eqs.~(\ref{IE},\ref{IO}) that the required integrands 
$F(\omega)$ have factors of the form $\sin^2 (Nxq)$ in the numerators and 
$\Delta_{1,N}$ in the denominators. 
In the limit $N \to \infty$ the factors $\sin^2 (N x q)$ in the numerators can be replaced by $1/2$.   
Next we note that the determinant $\Delta_{1,N}$ always has the 
following form:  
\bea
\Delta_{1,N}=A(q) \sin (Nq) + B(q) \cos (Nq)~,
\eea
where $A$ and $B$ are smooth complex-valued functions. 
We now obtain the following result for any  function 
$g(\theta, \phi)$ which is periodic in both variables:
\begin{eqnarray}
\lim_{N \rightarrow \infty} \int_{0}^{\pi} &d\theta& g(\theta, N \theta) = \lim_{N \rightarrow \infty} \f{1}{N} \int_{0}^{2 \pi (N/2)} d \phi g(\f{\phi}{N},\phi) \nn \\ &=& \lim_{N \rightarrow \infty} \f{1}{N}\sum_{i=1}^{i=(N/2)} \int_{2 \pi (i-1)}^{2 \pi i} d \phi g(\f{\phi}{N},\phi) = 
\lim_{N \rightarrow \infty} \f{1}{N}\sum_{i=1}^{i=(N/2)} \int_{0}^{2 \pi} d \psi g(\f{2 \pi (i-1)+ \psi}{N},\psi) \nn \\ &=& \lim_{N \rightarrow \infty} \sum_{i=1}^{i=(N/2)} \f{1}{N} \int_{0}^{2 \pi} d \psi g(\f{2 \pi (i-1)}{N},\psi) = \f{1}{2\pi}\int_{0}^{\pi} d \theta \int_{0}^{2\pi} d\psi g(\theta, \psi) \nn ~.
\end{eqnarray}
Using this we obtain:
\bea
\int_0^\pi dq \f{C(q)}{| A(q) \sin (Nq) + B(q) \cos (Nq)|^2} 
&=& \int_0^\pi C(q) dq \f{1}{2\pi} \int_0^{2 \pi}  d \psi \f{1}{| A(q) \sin \psi + B(q) \cos \psi|^2} \nn \\  
&=& \int_0^\pi dq  \f{C(q)}{| A(q) B^*(q) - A^*(q) B(q)| } 
\eea
Using this we  get the asymptotic forms of the  
various integrals in Eqs.~(\ref{IE},\ref{IO}),  and these lead to the results given in Eqs.~(\ref{int-even},\ref{int-odd}).
 When $\gamma_L =\gamma_R=\gamma$, we  can explicitly carry out the integrals appearing in these expressions and we get the following results.
 
{\bf Case (1)} - even $N$: 
\begin{eqnarray} 
I_{odd} &=& \f{m_b}{2} \f{2(1+\beta) + (\delta^2 +2\delta)2\beta + 2 \f{\beta^2}{1+2\beta \mu} \delta^2 (\delta+1)^2}{\sqrt{2 \beta +1}\sqrt{1+2\beta + 2\beta^2 \mu}}  - 4  \f{m_a}{1+2\beta \mu} \left( \sqrt{\f{1+2\beta +2\beta^2 \mu}{1+2\beta}} -1 \right)  \nn \\ 
&+&\f{m_b}{2} \f{|\delta| (\delta+1)^2}{\mu (1+2 \beta \mu) \sqrt{1+\delta^2}} +  m_a \f{2 \beta \mu}{1+2 \beta \mu} \left( 1- \f{|\delta|}{\sqrt{1+\delta^2}} \right)  \label{tempo-even} \\
I_{even} &=& m_a \left( 1 + \f{ - (1+\beta) + \beta( 2 \delta - \delta^2)  -\delta^2(1-\delta)^2 \f{\beta^2}{1+2 \beta \mu}}{\sqrt{1+2\beta}\sqrt{1+2\beta +2\beta^2 \mu}} + \beta \f{|\delta| (1-\delta)^2}{\sqrt{1+\delta^2}(1+2\beta\mu)} \right) \nn \\ 
&+&   \f{m_b}{1+2\beta \mu} \left( - \f{|\delta|}{\sqrt{1+\delta^2}} + \f{\sqrt{1+2\beta +2 \beta^2 \mu}}{\sqrt{1+ 2\beta}} \right)  
\label{tempe-even} \\
   J^{E} &=& \Delta T \f{\gamma}{\beta^2 \mu(1+4 \gamma^2)} [ 2 \beta +1 + 2\beta^2 \mu ( 1+\delta^2 -|\delta| \sqrt{1+\delta^2}) -\sqrt{(2\beta+1)(2\beta +1 +2 \beta^2 \mu)} ]  \nn \\   \label{curr-even}
\end{eqnarray}
where $\mu={2 m_a m_b}~, \delta = {m_a-m_b} ~,
\beta =\gamma^2 /(m_a m_b)$~.

{\bf Case (2)} - odd $N$: 
\bea
I_{odd} &=& I_{even}=\f{1}{2}~, \label{tempoe-odd} \\
J^{O} &=& \Delta T \f{\gamma B}{G^2} \left(1- \f{\sqrt{(F+H)^2 - G^2} + \sqrt{(F-H)^2 -G^2}}{2F}\right) \label{curr-odd}
\end{eqnarray}
where 
\begin{eqnarray}
    F &=& \f{B}{2C} (B^2 - 4 A C)^{1/2}~,~~  G= C \mu~,~~  
H= \f{B^2}{2C} - C(1-\mu) -A \nn \\ 
\text{and}~~ ~ A &=& \f{\delta}{m_b} - \f{\delta \beta}{m_a}~,~~ 
   B = \f{1}{m_b} + \f{2m_b \beta }{m_a}~,~~
   C = \f{\beta}{m_a}~. \nn
    \end{eqnarray}

%

\end{document}